%% file: commonality.tex
\documentclass[sigconf]{acmart}
\usepackage{fdiaz}
\usepackage{subcaption}

\AtBeginDocument{%
  }

\copyrightyear{2022}
\acmYear{2022}
\setcopyright{rightsretained}
\acmConference[RecSys '22]{Sixteenth ACM Conference on Recommender Systems}{September 18--23, 2022}{Seattle, WA, USA}
\acmBooktitle{Sixteenth ACM Conference on Recommender Systems (RecSys '22), September 18--23, 2022, Seattle, WA, USA}
\acmDOI{10.1145/3523227.3551476}
\acmISBN{978-1-4503-9278-5/22/09}

\input{notation.tex}

\begin{document}
\title[Measuring Commonality in Recommendation of Cultural Content]{Measuring Commonality in Recommendation of Cultural Content: Recommender Systems to Enhance Cultural Citizenship}

\author{Andres Ferraro}
\email{andresferraro@acm.org}
\orcid{0000-0003-1236-2503}
\affiliation{%
  \institution{McGill University}
  \city{Montr\'eal}
  \country{Canada}
}

\author{Gustavo Ferreira}
\email{gustavo.ferreira@mila.quebec}
\affiliation{%
  \institution{McGill University}
  \city{Montr\'eal}
  \country{Canada}
}

\author{Fernando Diaz}
\affiliation{
  \institution{Canadian CIFAR AI Chair}
  \country{}
}
\affiliation{%
  \institution{Google}
  \city{Montr\'eal}
  \country{Canada}
}
\email{diazf@acm.org}

\author{Georgina Born}
\affiliation{%
  \institution{University College London}
  \city{London}
  \country{United Kingdom}}
\email{g.born@ucl.ac.uk }

\input{abstract.tex}



\maketitle
\input{introduction.tex}
\input{motivation.tex}

\input{metrics.tex}
\input{methods.tex}

\input{results.tex}

\input{discussion.tex}
\begin{acks}
This research project forms part of the research program ‘Music and Artificial Intelligence: Building Critical Interdisciplinary Studies’ (MusAI), directed by PI Prof. Georgina Born, which has received funding from the European Research Council (ERC) under the European Union’s Horizon 2020 research and innovation programme (grant agreement No. 101019164). This project was supported in part by the Canada CIFAR AI Chairs program.
\end{acks}
\bibliographystyle{ACM-Reference-Format}
\bibliography{commonality.bib}
\end{document}

%% file: notation.tex
\newcommand{\ranking}[0]{\pi}
\newcommand{\rankings}[0]{\ranking}
\newcommand{\numusers}[0]{m}
\newcommand{\numitems}[0]{n}
\newcommand{\recall}[0]{\text{R}}
\newcommand{\user}[0]{u}
\newcommand{\users}[0]{\mathcal{U}}
\newcommand{\commonality}[0]{\text{C}}
\newcommand{\familiarity}[0]{F}

\newcommand{\catalog}[0]{\mathcal{D}}
\newcommand{\categories}[0]{\mathcal{G}}
\newcommand{\category}[0]{g}

%% file: abstract.tex
\begin{abstract}
Recommender systems have become the dominant means of curating cultural content, significantly influencing the nature of individual cultural experience. While the majority of academic and industrial research on recommender systems optimizes for personalized user experience, this paradigm does not capture the ways that recommender systems impact cultural experience in the aggregate, across populations of users. Although existing novelty, diversity, and fairness studies probe how recommender systems relate to the broader social role of cultural content, they do not adequately center culture as a core concept and challenge. In this work, we introduce commonality as a new measure of recommender systems that reflects the degree to which recommendations familiarize a given user population with specified categories of cultural content. Our proposed commonality metric responds to a set of arguments developed through an interdisciplinary dialogue between researchers in computer science and the social sciences and humanities. With reference to principles underpinning non-profit, public service media (PSM) systems in democratic societies, we identify universality of address and content diversity in the service of strengthening cultural citizenship as particularly relevant goals for recommender systems delivering cultural content. Taking diversity in movie recommendation as a case study in enhancing pluralistic cultural experience, we empirically compare the performance of recommendation algorithms using commonality and existing utility, diversity, novelty, and fairness metrics. Our results demonstrate that commonality captures a property of system behavior complementary to existing metrics and suggest the need for alternative, non-personalized interventions in recommender systems oriented to strengthening cultural citizenship across populations of users. In this way, commonality contributes to a growing body of scholarship developing `public good' rationales for digital media and machine learning systems.
\end{abstract}

%% file: introduction.tex
\section{Introduction}
\label{sec:introduction}

Online platforms that host cultural content (e.g. music, movies, and books) often use recommender systems to suggest and distribute items from their catalogs. These algorithmic recommendations are often designed to optimize personalization objectives such as precision or click-through rate \cite{gunawardana:receval-chapter} and, as such, are aligned with user-level metrics like retention and subscriptions.  

Academic and industry research on recommender systems has converged on personalization as a paradigm, and while metrics linked to personalization are productive, they do not capture the wider shaping effects of recommender systems in aggregate nor measure the effects of recommender systems across a population of users.  Research demonstrates that recommendations have cumulative effects in shaping the wider cultures and societies within which they are being used \cite{anderson2020}.  That said, these effects and how they inform recommender system evaluation and design remains relatively unexplored \cite{born:cifar}. 

We advocate for the design of recommender systems delivering cultural content to be motivated not only by personalization--or associated commercial interests--but also by appropriate normative principles. For guidance we turn to the normative principles underpinning non-profit, public service media (PSM) systems. We identify universality of address and content diversity in the service of strengthening cultural citizenship as particularly relevant for recommender systems delivering cultural content. If personalization attempts to maximize individual user satisfaction with a platform, then these PSM principles aim more to enhance the commonality of diverse cultural experiences across a population, building cultural citizenship. In this way, we contribute to a growing body of scholarship developing `public good' rationales for digital media and machine learning systems~\cite{murdock2005building,andrejevic2013public,van2018public,moe2008dissemination,born:21stcentury,born:ecology,unterberger2021public}

In this light, in this paper we propose and discuss a new evaluation metric that measures the degrees to which a system familiarizes a given user population with a specified category or group of categories of cultural content.  Our proposed commonality metric responds to a set of arguments developed through an interdisciplinary dialogue between researchers in computer science and the social sciences and humanities.  

As a case study in the application of these normative principles and their goal of enhancing cultural citizenship across a population, we consider movie recommendations. We empirically compare the performance of more than twenty recommendation algorithms using the proposed commonality metric with existing utility, diversity, novelty, and fairness metrics. We analyze how our commonality metric complements previous metrics and discuss how it is aligned with the intended purpose. Our aim with this work is to measure commonality in the consumption of diverse categories of movies that are generally under-represented by existing recommender systems. To date, criticisms of recommender systems and machine learning systems for their capacity to reproduce forms of bias and discrimination have been met by interventions designed to redress such biases at the level of individual users. However, it is also important to identify means of counteracting biases and enhancing diversity as common experiences across a population of users – by developing recommender systems that promote diversity through counteracting racism, sexism, and the neglect of non-Western content as common experiences. Only in this way can the wider cultural changes called for by anti-racist and feminist critics as well as those sympathetic to these criticisms from the RecSys community be delivered.

%% file: motivation.tex
\section{Background}\label{sec:motivation}


Recommender systems have become the dominant means of curating cultural content in the digital era. Curation – or the selection and promotion of content to be distributed to consumers – is, however, a historical constant: consumers have always encountered the cultural content they wish to consume via some type of curation. Today, when cultural curation supplied by recommender systems is multiplied across billions of recommendations presented to users by online platforms, they significantly influence the nature of individual cultural experiences \cite{tomlein2021audit}. But this influence is also magnified and multiplied across time and across populations, regions, and cultures. In the short term, recommender systems influence individual cultural consumption and taste. In the medium and long terms, by employing data on consumer behavior and influencing consumer choices, they can shape cultural literacies as well as population-wide trends in consumption and taste \cite{born:cifar}. As a result, unlike previous means of distribution, there is a high degree of automatized intervention in the way people and communities encounter and experience cultural content. Despite their personalized address, recommender systems therefore have cumulative effects in shaping the wider cultures and societies within which they are being used. These effects have been relatively unexplored as a focus of research in the RecSys community.


Academic and industrial research on recommender systems has converged on personalization as a paradigm. While metrics linked to personalization and `user relevance' are productive (e.g., NDCG, precision, recall), they do not capture the wider, aggregate shaping effects of recommender systems on patterns of cultural consumption, taste and literacy as described above.  That is, they do not measure the effects of recommender systems' use across populations of users. 

Although concerned with and sensitive to the broader social role of cultural content, existing diversity and fairness evaluations of recommender systems do not adequately support the rich set of goals system designers might have. Typically, diversity metrics are limited to the goal of capturing the variety of content offered \textit{within} a recommendation list; they may consider categorizations of the content, distances in a latent space, or simply how many different items are recommended \cite{schedl2015, Yucheng2018, anderson2020, Han2017ASO}. Aligned with the goals of personalization, the formulation of these diversity metrics sometimes optionally considers the relevance of the content for users, assuming that what a user consumed in the past indicates what they are still interested in, so recommendations should be limited to such categories. While related novelty metrics measure the newness of items or categories of recommendations, they are still individualized and agnostic about \textit{what type} of content is new to the user. The work on fairness addresses specific topics of increasing biases and under-representation of particular groups \cite{Ekstrand2022}.  \textit{Provider fairness} metrics typically consider how many different groups of content providers appear in recommendations and assume a given distribution that it is desired to match. \textit{Consumer fairness} considers disparate treatments of the system to different groups of consumers.  Recent research proposed more general \textit{multi-stakeholder fairness}, acknowledging the impact recommender systems have for the different groups of individuals \cite{milano2021ethical,burke2016towards,sonboli2022}.



We suggest that it is timely for the design of recommender systems delivering cultural content to be motivated not only by individualized interests but by appropriate normative principles oriented to furthering the democratic development of contemporary societies. By normative we refer to principles considered to provide models of morally, ethically and/or politically right or just action or behavior in the interests of democratic societies as well as individuals. In this way we contribute to growing scholarship developing `public good' rationales for digital media and ML systems~\cite{murdock2005building,andrejevic2013public,van2018public,moe2008dissemination,born:21stcentury,born:ecology,unterberger2021public}, advocating `a computational politics wedded to emancipation and human flourishing'~\cite{stark2018algorithmic}. For guidance we turn to the principles underpinning non-profit, public service media (PSM) systems, proposing that `a public service rationale is as pertinent as ever in the digital era'~\cite{andrejevic2013public}. The normative ideas underpinning public service media developed over the last century in the context of democratic states committed to enhancing democratic and representative channels of communication.

A substantial body of research in media and political theory has identified the normative principles informing PSM systems, among them universality, citizenship, and diversity~\cite{scannell1989public,raboy1996public,born:21stcentury,born:ecology,born-prosser:psm-principles,born:2006digitilizing}. We consider this triad – universality (or commonality) of address, citizenship, and diversity of content – as particularly relevant for recommender systems delivering cultural content, since together they answer calls for digital media systems to enhance cultural citizenship~\cite{born:2006digitilizing}. The concept of cultural citizenship has become foundational for democratic political theories in the last two decades. It responds to recognition of the challenges posed by globalization, migration, the growing heterogeneity of the populations of nation states, and the intensification of identity politics among subaltern and marginalized groups~\cite{rosaldo1994cultural,miller2001introducing}. It draws attention to a `new domain of cultural rights [involving] the right to symbolic presence, dignifying representation', and `the maintenance and propagation of distinct cultural identities'~\cite{pakulski1997cultural}. In this light, PSM – and other democratic distribution and curation media – should promote cultural citizenship by curating and disseminating a plurality of cultural content stimulating intercultural dialogue and `acceptance of, and respect for, cultural diversity'. In this way PSM and other democratic media can act both as a force `for social cohesion and integration' and as a forum for pluralistic cultural experience among groups and communities coexisting in democratic societies~\cite{jakubowicz2007public}. Both universality or commonality--the provision of common cultural experiences--and diversity of content are therefore essential to the strengthening of cultural citizenship: `mutual cultural recognition and the expansion of cultural referents… are dynamics essential to the well-being of pluralist societies. But this does not obviate the need also for integration--for the provision of common [cultural] experience and the fostering of common identities'~\cite{born:2006digitilizing}. Scholarship on these matters emphasizes that implementing principles like universality (commonality), citizenship, and diversity require `alternative success metrics… focused on PSM's impact on democracy and the public sphere' which address users `as citizens and not just… as consumers'~\cite{unterberger2021public}. Such metrics will enable democratic media to adapt to the present by advancing `cultural citizenship and the needs of the digital society'~\cite{jakubowicz2007public}.


As a case study we take diversity in movie recommendation, with the aim of measuring commonality in the consumption of diverse categories of movies that are as yet under-represented in existing recommender systems. Current recommender systems are criticized for a tendency to reproduce or exacerbate wider forms of cultural and social discrimination~\cite{olteanu2019,ekstrand2018}. Previous works show that movie recommender systems may generate feedback loops between recommendation and consumption. As a result, they amplify popularity bias, promote narrow options in terms of cultural content, and homogenize users' identity profiles, with a stronger effect on gender minorities\cite{mansoury2020feedback}. In movie and music domains, commercial popular content may be privileged for users in dominant groups according to age and gender~\cite{ekstrand2018all}. From the provider side in the music domain, female and non-binary artists can be under-represented in recommendations, reflecting industry gender biases and reducing the diversity of content being recommended which may affect users' future streams~\cite{epps2020artist, ferraro2021break}. These provider biases shown by under-representation of cultural content with respect to gender, race, class, and region correspond to `core-periphery dynamics and geographical inequality' in the cultural industries~\cite{tofalvy2021splendid,camposinequalities,verhoeven2019re}. Recommender systems often mirror these inequalities, promoting Western-centric popular cultural content in the English language, released by major producers~\cite{tofalvy2021splendid,voit2021bias}. Generally, criticisms of recommender systems and machine learning systems for reproducing such forms of bias and discrimination have been met by personalized recommender systems interventions aimed at redressing bias at the level of individual users. In light of our discussion of culture, we contend that it is also important to identify means of counteracting bias and enhancing the diversity of content offered across a population of users. The proposed commonality metric achieves this by measuring common experiences of diversity at the aggregate level. Assuming a democratic media environment, the metric provides a way of assessing whether recommender systems are contributing to the strengthening of cultural citizenship by systematically promoting diversity within a given type of cultural content (here, movies). At the same time, it has the potential to assist in counteracting racism, sexism, and the neglect of non-Western and non-mainstream content across a user population (via commonality). In this way, the metric is a means of measuring the extent of the kinds of wider cultural changes called for by anti-racist and feminist critics as well as by those sympathetic to criticisms of existing recommender systems.


%% file: metrics.tex
\section{Measuring Commonality}
\label{sec:commonality}
Recall that we are interested in measuring the extent to which users will gain a shared familiarity with a set of promoted categories. The promoted categories, we suggest, will be identified and curated by experts in a relevant field (here, movies) seeking to promote a plurality of cultural content in the service of strengthening cultural citizenship.  We contrast this with \citeauthor{mehrotra:fair-marketplace}'s purely statistical method for selecting under-represented categories \cite{mehrotra:fair-marketplace}, which may surface under-represented content misaligned  with our twin goals of boosting a diversity of cultural experience across a user population, and at the same time enhancing their common experience of that diversity. An expert may opt to promote, for example, movies by female directors as well as those produced for non-Western markets.

The  \textit{commonality} of a system captures the probability that \textit{every} user simultaneously gains familiarity with the editorially-selected categories. Let $\users$ be the set of system users where $\numusers=|\users|$ and $\catalog$ the catalog of items where $\numitems=|\catalog|$.  Given a user $\user\in\users$,  a system produces a ranking $\ranking_\user$ of items in $\catalog$; we will use $\ranking$ to refer to the set of all rankings.  We are interested in measuring the degree to which the system supports the normative value of commonality.  Given a set of editorially-selected categories $\categories$, we can compute the commonality of a system with respect to a single category $\category\in\categories$ as the probability that every user has become familiar with $\category$ under the system's ranking,
\begin{align}
    \commonality_\category(\rankings)&=p(\familiarity_{1,\category},\ldots,\familiarity_{\numusers,\category}|\rankings)\nonumber\\
    &=\prod_{\user\in\users}p(\familiarity_{\user,\category}|\ranking_\user)
\end{align}
where $\familiarity_{\user,\category}$ is a binary random variable indicating the familiarity that user $\user$ has with category $\category$. 

In order to estimate $p(\familiarity_{\user,\category}|\ranking_\user)$, the familiarity of a user with a category after the recommender intervention, we adopt a standard browsing model from existing evaluation metrics.  Specifically, we are interested in the comprehensiveness of a user's exposure to items from $\category$ after browsing the ranking,
\begin{align}
    p(\familiarity_{\user,\category}|\ranking_\user)&=\sum_{i=1}^{\numitems}p(i)\recall(\ranking_\user,i,\category)
\end{align}
where $p(i)$ is the probability that the user stops at rank $i$ and $\recall(\ranking,k,\category)$ is the recall of items with category $\category$ in $\ranking$ at rank cutoff $k$.  For our experiments, we adopt an exponential discount, $p(i)=(1-\gamma)\gamma^{i-1}$ based on the rank-biased precision browsing model \cite{moffat:rbp}.  
For a set of categories, we measure commonality using the mean commonality, $\commonality_\categories\left(\rankings\right)=\mathbb{E}_{\category\in\categories}\left[\commonality_\category(\rankings)\right]$.


\if false
\begin{align*}
p(u \text{ familiar with }c) &= \sum_{i=1}^{|\catalog|} p(u\text{ familiar with }c, u \text{ stops at }i) \\
&= \sum_{i=1}^{|\catalog|} p(f, i) \\
&= \sum_{i=1}^{|\catalog|} \sum_{j=0}^{m} p(f, i, u \text{ sees } j  \text{ category items})\\
&= \sum_{i=1}^{|\catalog|} \sum_{j=0}^{m} p(f|i, u \text{ sees } j  \text{ category items})p(u \text{ sees } j  \text{ category items}|i)p(i)\\
&= \sum_{i=1}^{|\catalog|} p(i) \sum_{j=0}^{m} p(f|u \text{ sees } j  \text{ category items})p(u \text{ sees } j  \text{ category items}|i)\\
&= \sum_{i=1}^{|\catalog|} p(i) p(f|m_{\pi_{[1,i]}})\\
\end{align*}
\fi
%

\if false
\begin{align*}
log( \prod_\user \sum_{i=1}^{|\catalog|} p(i) p(f|m_{\pi_{[1,i]}}) &= \sum_\user log( \sum_{i=1}^{|\catalog|} p(i) p(f|m_{\pi_{[1,i]}}))\\
&= \sum_\user log( \sum_{i=1}^{|\catalog|} exp( log(p(i) p(f|m_{\pi_{[1,i]}})))) \\
&= \sum_\user log( \sum_{i=1}^{|\catalog|} exp( a_i)) \\
&= \sum_\user log( \sum_{i=1}^{|\catalog|} exp( a_i - b_u)) + b_u \\ 
\end{align*}
\fi

%% file: methods.tex
\section{Experiments}
\label{sec:experimetns}

The goal of these experiments is to empirically compare commonality with existing metrics when ranking different recommender systems. Using movies as a case study, we focus on analyzing how commonality complements previous metrics in the consumption of diverse categories of movies that are generally under-represented by existing recommender systems. We then discuss how commonality is aligned with the intended purpose of promoting a shared plurality of cultural content in the service of strengthening cultural citizenship.



\underline{Data}: 
We use the movielens-1m dataset, which contains 1,000,209 ratings of ~3,900 movies from 6,040 users from the movielens platform. Using a separate dataset\footnote{https://www.themoviedb.org/}, we augmented the movielens movies with metadata including country of production, gender of the director, original language, and keywords collected from the movie's description.  For this dataset, we used rankings from multiple recommendation systems prepared by \citet{valcarce2018}. Following the method described by the authors we converted to binary relevance labels considering ratings of 4 and 5 as relevant.

\underline{Categories}: 
We selected categories of movies that are typically under-represented by movie 
recommender systems. Specifically, we consider female directors (under-representation by gender); independent film (under-representation by industry sector); and several sources of non-Western film (under-representation by geographical and linguistic inequality).  We use categorical gender data, acknowledging the limitations of this framing \cite{hoffmann:fairness-failure}.  For geographic categories, we use the country of production for the following regions of the world: 
South America, Central America, North Africa, South Africa, West Africa, Mid Africa, Southeast Asia, South Asia, Western Asia, Central Asia and East Asia.
We consider, broadly, non-English language movies as a separate category.  
And, finally, we use keywords to create categories with selected movies whose categories contain ``independent films'', ``LGBT'', and ``transgender''. We manually checked whether these keywords can be trusted to represent specific identities. 

\underline{Baseline Metrics}: 
We compare commonality against three classes of metrics: utility metrics, diversity metrics, and fairness metrics \cite{canamares:offline-recsys-eval}.  For utility, we consider normalized discounted cumulative gain (NDCG) and reciprocal rank (RR).  For diversity, we consider $\alpha$-NDCG and intent-aware expected reciprocal rank (ERR-IA). For fairness, we use ranking-based statistical parity (RSP)  and ranking-based equal-opportunity (REO) \cite{zhu2020measuring}.




%% file: results.tex
\section{Results}\label{sec:results}

We were first interested in understanding the redundancy between commonality and existing metrics. To this end, we measured Kendall's $\tau$ between system rankings according to commonality with system rankings according to our baseline metrics.
We present the results of this analysis in Table \ref{tag:metric-correlation}.  As expected, the correlation between commonality and utility metrics (NDCG, RR) is negative and significant, indicating that we would tradeoff commonality and utility (personalization) if selecting from existing systems.  Moreover, this suggests that, for existing systems, if we select the higher utility systems during model development, it will compromise our goals of commonality and cultural citizenship.  We observed a similar negative correlation with existing diversity metrics, largely due to the utility-orientation of the metrics.  Finally, we did not find evidence of correlation between commonality and fairness metrics.  Collectively, these results suggest that commonality measures properties of a set of recommendations not present in existing metrics.   


\begin{table}
  \caption{Correlation between commonality with existing metrics.  Kendall's $\tau$ between rankings of runs. **~indicates p<0.05.}\label{tag:metric-correlation}
  \centering
          \centering
          \begin{tabular}{cc|cc|cc}
            \hline

            NDCG&
            RR&
$\alpha$-NDCG & 
ERR-IA &
REO & 
RSP \\
              \hline
              -0.4476** & 
              -0.5048** & 
              -0.5333**& 
              -0.5143** & 
              0.2316&
              0.1053 \\
               
\hline
  \end{tabular}
\end{table}

\begin{figure*}
\centering
\begin{subfigure}{.43\textwidth}
  \centering
  \includegraphics[width=0.95\textwidth]{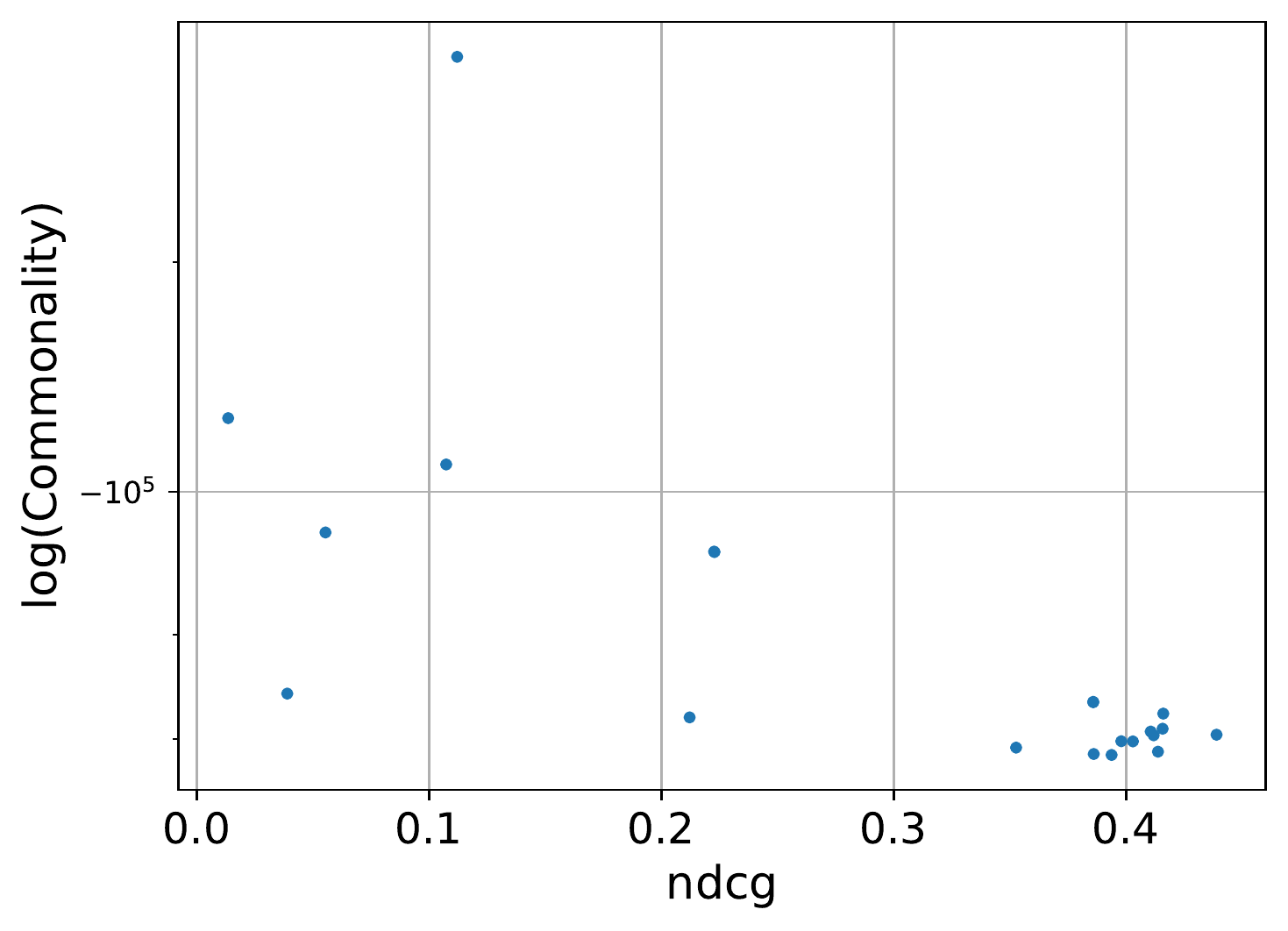}
   \caption{utility vs commonality}
   \Description[commonality vs ndcg]{Commonality compared with ndcg for multiple recommender systems shows a negative correlation.}
    \label{fig:ncdg-comm}
\end{subfigure}%
\begin{subfigure}{.58\textwidth}
  \centering
  \includegraphics[width=0.98\textwidth]{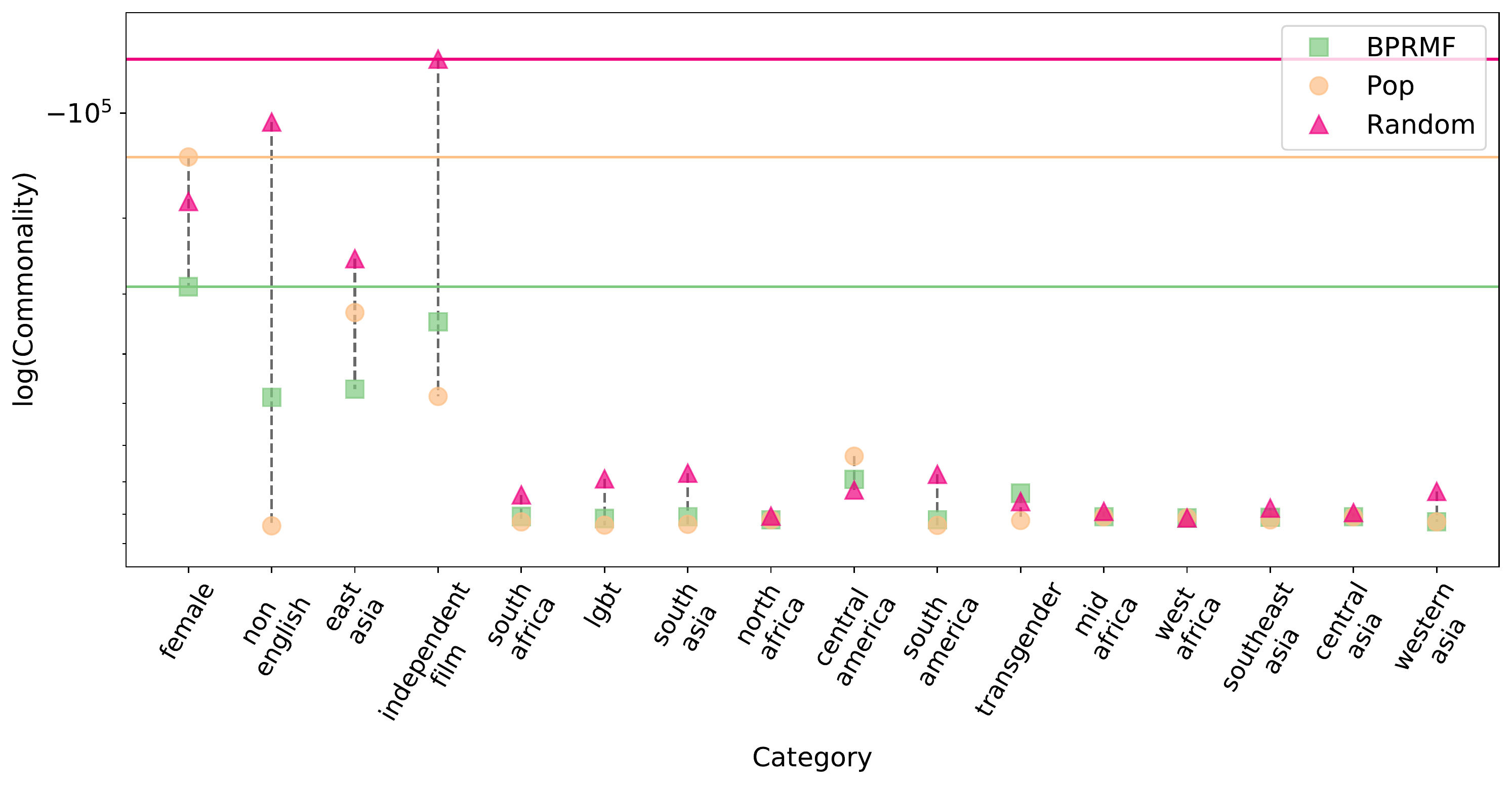}
    \caption{disaggregated commonality}
    \Description[disaggregated commonality]{Differences in commonality per category for recommendations based on random, popularity, and BPRMF models compared to the mean values. Pop recommender gives higher commonality for female director films while giving lower overall performance for commonality.}
    \label{fig:comm-cats}
\end{subfigure}
\caption{Behavior of commonality.  \ref{fig:ncdg-comm} Scatterplot of mean NDCG versus mean commonality across categories.  Each point is a single recommender system run.  \ref{fig:comm-cats} Per-category commonality values for recommendations based on  random, popularity, and BPRMF models.  Horizontal lines indicate the mean commonality across categories.   }
\end{figure*}

In order to explore the relationship between commonality and utility, we plotted the NDCG and commonality values for our twenty runs in Figure  \ref{fig:ncdg-comm}.   The run with the highest NDCG, Bayesian personalized ranking matrix factorization (BPRMF), also exhibited a very low average commonality across categories, relative to other runs; we contrast this with the run with the lowest NDCG, random, that exhibited comparatively much higher average commonality across categories.  The run with the highest average commonality, SVD, on the other hand, generated lower-utility rankings.   



In Figure \ref{fig:comm-cats}, we further disaggregate the metric for three runs: BPRMF, random, and popularity. We observed that popularity-based ranking offers the highest commonality for the ``female'' and ``north africa'' categories while random ranking offers a the best commonality for all categories except ``south america''.  In general, personalization-based recommendation had low commonality across all categories. 

%% file: discussion.tex
\section{Conclusions and Future Work}
\label{sec:discussion}

In this work, motivated by defining metrics for recommendation of cultural content, we developed a method to measure alignment with principles of cultural citizenship that we adapted from the PSM literature.

Our proposed commonality metric emphasizes shared familiarity, by which is intended the simultaneous exposure of users to content from selected categories.  This definition, captured by the joint probability of familiarity events, is worth exploring for both theoretical and pragmatic reasons.  Theoretically, our approach to shared exposure is very conservative, penalizing the metric if even a single user has a low probability of exposure.  That said, this conservativeness attenuates systematic under-exposure to certain categories.  Pragmatically, since the joint probability is often quite small for our evaluated systems, we can encounter numerical stability issues.  This is especially salient if we average commonality across categories, where a single category may have an exponentially larger value and dominate the mean.  However, this may indicate gross and systemic under-performance of existing systems in terms of commonality.  

In addition to commonality, we introduce a relatively simple model of familiarity based on recall.  We believe there is opportunity to develop alternative models of familiarity that consider a user's previous experience with the category or other contextual information.  However, the design of a familiarity model should be aligned with the concept of shared experience, meaning that, even if a user has engaged with content from a category in the past, \textit{re-exposing} them may promote commonality at the risk of over-satiating users with niche interests, a topic of recent research \cite{leqi:satiation}.  

Our results demonstrate that existing high-utility recommendation algorithms under-perform in terms of commonality.  We believe that exploring the space of commonality-informed recommendation can produce algorithms that perform substantially better in terms of commonality while maintaining high utility.

In summary, our preliminary results indicate a new class of evaluation metrics specifically aimed at measuring the effects of recommendation systems in terms of common exposure to diverse items within a given area of cultural content. Our future work includes assessing important evaluative properties of commonality such as generalizability and robustness. We are also interested in the joint optimization of both personalization and commonality as a new class of algorithm. Future extensions to our commonality metric may also consider the cumulative effects over time of recommender systems attuned to enhancing common experience of diverse content and thereby cultural citizenship.
